\documentclass[aps, preprint, superscriptaddress]{revtex4-1}
\usepackage{amsmath,amsfonts,graphicx,epstopdf,hyperref}
\usepackage{natbib}

\begin{document}
\abovedisplayskip=20pt
\abovedisplayshortskip=20pt
\belowdisplayskip=20pt
\belowdisplayshortskip=20pt

\title{Suboscillations with arbitrary shape}

\author{Ioannis Chremmos}
\email{jochremm@central.ntua.gr}
\address{Hellenic Electricity Distribution Network Operator S.A.}
\address{School of Electrical and Computer Engineering, National Technical University, GR 157-73, Athens, Greece}

\author{Yujie Chen}
\address{State Key Laboratory of Optoelectronic Materials and Technologies, School of Electronics and Information Technology, Sun Yat-sen University, Guangzhou 510275, China}

\author{George Fikioris}
\email{gfiki@ece.ntua.gr}
\address{School of Electrical and Computer Engineering, National Technical University, GR 157-73, Athens, Greece}

\date{\today}

\begin{abstract}
We report a method for constructing bandpass functions that approximate a given analytic function with arbitrary accuracy over a finite interval. A corollary is that bandpass functions can be obtained that oscillate arbitrarily slower than their minimum frequency component, a counter-intuitive phenomenon known as \textit{suboscillations}. 
\end{abstract}
\maketitle
\section{Introduction}
It is known that a bandlimited function (a function whose spectrum is supported in a band of frequencies of the form $|\omega| \leq \omega_{\text{max}}$) can oscillate arbitrarily faster than its maximum frequency component and that it can do so over arbitrarily long, yet finite, intervals. Early manifestations of this counter-intuitive phenomenon can be traced back to concepts such as superdirective electromagnetic radiation \cite{Oseen1922} and optical imaging beyond the diffraction limit \cite{Toraldo1952}, while more explicit studies first appeared within the field of information theory and signal processing \cite{Bond1958, Bucklew1985}. The term \textit{superoscillations} was actually coined as part of a reinvention of the phenomenon within quantum mechanics \cite{Aharonov1990TimeTranslation, Berry1994Billiards}, and was also adopted within optics where superoscillations are exploited towards superresolution imaging \cite{BerrySuperOscillations, Rogers2013}.

A complementary to superoscillations phenomenon has recently been introduced under the term \textit{suboscillations} and was used to demonstrate super defocusing of a light beam \cite{Eliezer_Suboscillations}. Suboscillations refer to the ability of a bandpass function (a function whose spectrum is supported in a two-sided band of frequencies of the form $\omega_{\text{min}} \leq |\omega| \leq \omega_{\text{max}}$ with $\omega_{\text{min}} > 0$) to oscillate arbitrarily slower than its minimum frequency component $\omega_{\text{min}}$. The existence of such functions was demonstrated in \cite{Eliezer_Suboscillations} using the Fourier-type integral representation of \cite{BerryFasterThanFourier}, where the suboscillations are due to the presence in the integrand of a Dirac-function-like factor centered in the complex plane. Although it provides examples of suboscillatory functions, this approach offers little flexibility in designing the actual shape of the function in the suboscillatory region.

In this brief communication we show that it is possible to construct bandpass functions whose shape over a given interval approximates a desired analytic profile with arbitrary accuracy. More specifically, given an (arbitrarily small) $\epsilon>0$ and an analytic function $s(t)$ (analytic for all values of the real ``time" variable $t$) and an (arbitrarily long) interval $(a,b)$, we show how to construct a function $f_{\text{BP}}(t)$ whose spectrum is strictly zero outside the two-sided frequency band $\omega_{\text{min}} \leq |\omega|  \leq\omega_{\text{max}}$ and which satisfies 
\begin{equation}
|f_{\text{BP}}(t) - s(t)| < \epsilon 
\label{eq:approximation}
\end{equation}
for $t \in (a,b)$. Although an arbitrary analytic $s(t)$ is generally assumed, the case of a $s(t)$ that varies (or oscillates) at time scales larger than $\omega^{-1}_{\text{min}}$ is particularly interesting, since in this case the corresponding bandpass function $f_{\text{BP}}(t)$ oscillates in the interval $(a,b)$ slower than its minimum frequency component $\omega_{\text{min}}$, which is the counter-intuitive fact about suboscillations. Even more surprising is perhaps the extreme case of a constant $s(t)$ as it leads to functions that stay (almost) constant over an arbitrarily long internal without containing a dc component ($\omega=0$) in their spectrum.

\section{The method}
In \cite{Chremmos_SOPolynomial} we showed that it possible to construct bandlimited functions (functions whose spectrum is supported in a frequency band of the form $|\omega| \leq \omega_{\text{max}}$) that approximate a given polynomial $p_N(t)$ ($N$ being the degree) with arbitrarily small error over an arbitrarily long, yet finite, time interval. Such a function becomes superoscillatory in this interval when the corresponding polynomial oscillates faster than the maximum frequency component $\omega_{\text{max}}$. These functions were obtained as the product of the polynomial $p_N(t)$ with a bandlimited function $e(t)$ (which we termed \textit{envelope}) whose Fourier transform has at least $N-1$ continuous derivatives (namely belongs to $C^{N-1}(-\infty,\infty)$) and a $N^\text{th}$ derivative of bounded variation. The envelope has to be close to unity and sufficiently flat in the considered interval so that the deviation of the product $p_N(t)e(t)$ from the desired profile $p_N(t)$ stays smaller than a given $\epsilon > 0$ for all $t$ in the considered interval.

Using the method of \cite{Chremmos_SOPolynomial}, we obtain a bandpass function $f_{\text{BP}}(t)$ that satisfies \eqref{eq:approximation} as follows: We first construct a bandlimited function $f_{\text{BL}}(t)$ that approximates the analytic function $s(t)e^{-i \Omega t}$ in the interval $(a,b)$, where $\Omega > 0$. This is done by letting $p_N(t)$ be a polynomial (e.g. a Taylor polynomial) that approximates $s(t)e^{-i \Omega t}$ with accuracy $\epsilon_1$ in the interval $(a,b)$, and subsequently by designing $f_{\text{BL}}(t)$ so as to approximate $p_N(t)$ with accuracy $\epsilon_2$ in the same interval. One then has by the triangle inequality
\begin{equation}
\left|f_{\text{BL}}(t) - s(t)e^{-i \Omega t}\right| \leq \left|f_{\text{BL}}(t) - p_N(t) \right| + \left|p_N(t) - s(t)e^{-i \Omega t}\right| < \epsilon_2 + \epsilon_1, 
\label{eq:triangle}
\end{equation}
for $t \in (a,b)$. As we illustrate in Section \ref{sec:Example}, the accuracies $\epsilon_2$ and $\epsilon_1$ can be tuned independently, the first through the flatness of the envelope function $e(t)$ in the interval $(a,b)$ (by means of the dilation factor discussed in \cite{Chremmos_SOPolynomial}), and the second through the degree $N$ of the Taylor polynomial $p_N(t)$, so that the sum $\epsilon_2 + \epsilon_1$ can eventually be made smaller than any given $\epsilon > 0$. We then obtain from \eqref{eq:triangle}
\begin{equation}
\left|f_{\text{BL}}(t) - s(t)e^{-i \Omega t}\right| = \left|f_{\text{BL}}(t)e^{i \Omega t} - s(t)\right| < \epsilon, 
\label{eq:f_sup}
\end{equation}
for $t \in (a,b)$, where $f_{\text{BL}}(t) = p_N(t)e(t)$.

Now, without loss of generality, we assume that the envelope function $e(t)$ belongs to the Paley-Wiener space $\text{PW}_{\pi}$ of square integrable functions whose Fourier transform is supported in $[-\pi,\pi]$, namely the space of bandlimited functions with finite energy and bandwidth $\pi$. The function  $f_{\text{BL}}(t) = p_N(t)e(t)$ is also a member of this space since its Fourier transform is equal to a finite sum of derivatives of the Fourier transform of $e(t)$ up to order $N$ which are also members of $\text{PW}_{\pi}$ \cite{Chremmos_SOPolynomial}. It follows that $f_{\text{BL}}(t)e^{i \Omega t}$ is a finite-energy function whose spectrum is supported in $[\Omega-\pi, \Omega+\pi]$. For $\Omega > \pi$ this frequency band does not contain zero hence this function is by definition bandpass. It is finally obvious from the inequalities \eqref{eq:approximation} and \eqref{eq:f_sup} that the function 
\begin{equation}
f_{\text{BP}}(t) = f_{\text{BL}}(t) e^{i \Omega t} = p_N(t) e(t) e^{i \Omega t}. 
\label{eq:result}
\end{equation}
satisfies our requirements. The analytic function $s(t)$ must of course oscillate in the interval $(a,b)$ slower than the minimum frequency component $\omega_\text{min}=\Omega-\pi$ (for example $s(t) = \sin(\omega_l t)$ with $\omega_l < \omega_{\text{min}}$) if the bandpass function $f_{\text{BP}}(t)$ is aimed to be suboscillatory.

Note that the obtained bandpass function has its spectrum confined in the positive frequency band $[\Omega-\pi, \Omega+\pi]$ with no components in the corresponding negative frequency band $[-\Omega-\pi,\pi-\Omega]$. Suboscillatory functions with both positive and negative frequency components can be obtained through the following general recipe: The function $s(t)$ is arbitrarily split into two analytic parts as $s(t) = s_+(t) + s_-(t)$. We then use the method of \cite{Chremmos_SOPolynomial} to construct two bandlimited functions $f_{\text{BL}}^+(t)$ and  $f_{\text{BL}}^-(t)$ (with the same bandwidth $\pi$) that approximate the analytic functions $s_+(t)e^{-i\Omega t}$ and $s_-(t)e^{i\Omega t}$, respectively, in the interval $(a,b)$. It immediately follows that the function
\begin{equation}
f_{\text{BP}}(t) = f_{\text{BL}}^+(t)e^{i\Omega t} + f_{\text{BL}}^-(t)e^{-i\Omega t} 
\label{eq:two-side}
\end{equation}
is a bandpass function that approximates $s(t)$ in the interval $(a,b)$ and has the two-sided Fourier transform $F_{\text{BL}}^+(\omega-\Omega) + F_{\text{BL}}^-(\omega+\Omega)$, where $F_{\text{BL}}^+(\omega)$ and $F_{\text{BL}}^-(\omega)$ are, respectively, the Fourier transforms of the two bandlimited functions. As mentioned, the splitting of $s(t)$ into two terms is arbitrary. Equation \eqref{eq:result} corresponds to $s_+(t) = s(t)$ and $s_-(t) = 0$. Another choice could be $s_+(t) = s_-(t) = s(t)/2$.

\section{Example}
\label{sec:Example}
\begin{figure}[t]
\includegraphics[width=1.0\textwidth]{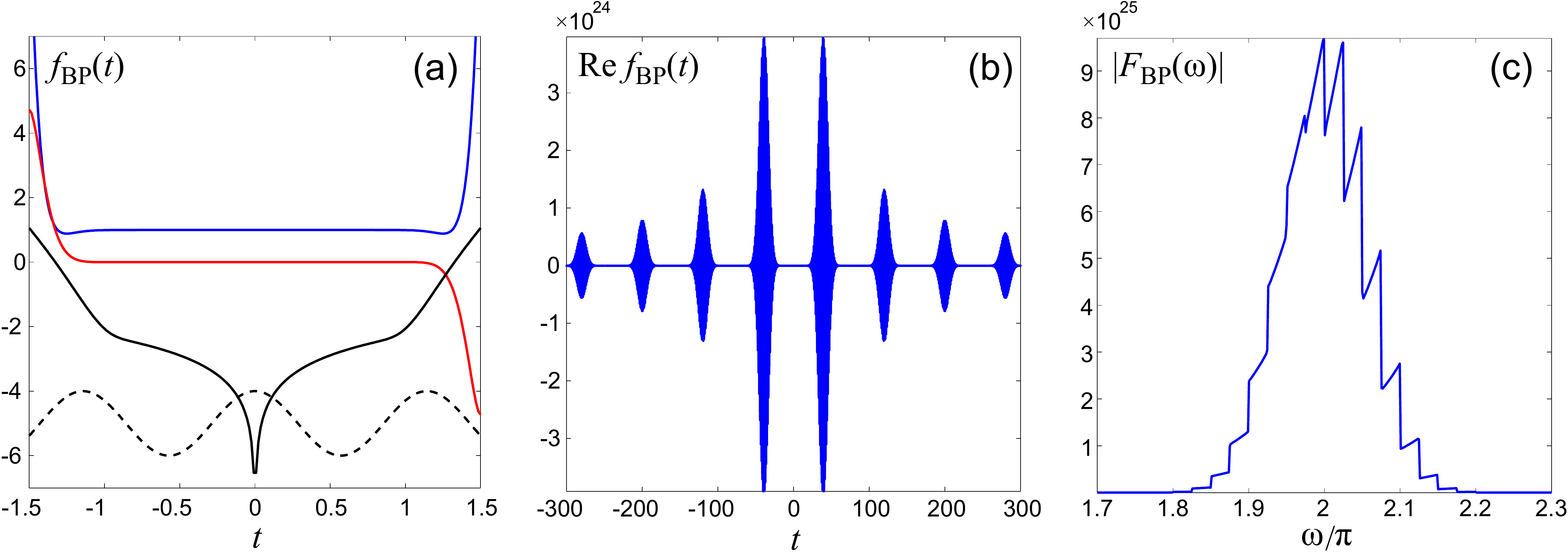}
\caption{The suboscillatory function $f_{\text{BP}}(t) = e^{i 2 \pi t} \text{sinc} ^{20} \left( \frac{t}{80} \right) \sum _{n=0}^{19} \frac{\left(-i 2 \pi t\right)^n}{n!}$. (a) Real (blue line) and imaginary (red line) part of $f_{\text{BP}}(t)$, zooming in the suboscillatory interval $(-1,1)$. The solid black line is $\text{log}_{10}|f_{\text{BP}}(t) - 1|$. The dashed black line is $\text{cos}(7 \pi t /4) - 5$ (the 5-unit shift is merely for illustration purposes). (b) Real part of $f_{\text{BP}}(t)$ in an extended interval. The oscillations of the modulation factor $e^{i 2\pi t}$ are too dense to be distinguished. (c) Magnitude of the Fourier transform of $f_{\text{BP}}(t)$. The vertical segments indicate points of discontinuity.}
\label{fig:figure_1}
\end{figure}

As an example, let us construct a suboscillatory function whose spectrum is zero outside the frequency band $[\Omega-\pi, \Omega+\pi]$, $\Omega > \pi$, and which approximates the constant function $s(t)=1$ in the interval $(-a,a)$. Since the interval contains $t=0$, we can use the Maclaurin series expansion
\begin{equation}
s(t) e^{ - i \Omega t} = e^{ - i \Omega t} = \sum _{n=0}^{N} \frac{\left(-i \Omega t\right)^n}{n!} + R_N(t), 
\label{eq:Maclaurin}
\end{equation}
where the first term is interpreted as the polynomial $p_N(t)$ of the previous discussion, and $R_N(t)$ is the remainder. A corollary of Taylor's theorem (\cite{Ahlfors_Complex}, p. 179) is that the remainder is bounded according to
\begin{equation}
|R_N(t)| \leq \left( \frac{\Omega t}{r} \right)^{N+1} \frac{e^r}{1 - \Omega t / r},
\label{eq:Remainder}
\end{equation}
where $r > \Omega t$ can be chosen freely because the complex function $e^z$ is analytic over the entire complex plane. The formula \eqref{eq:Remainder} shows that, for any $\epsilon_1>0$, a $N^*$ exists that ensures $|R_{N*}(t)|<\epsilon_1$ for $N>N^*$ and $|t|<a$. Having determined an appropriate order of the Taylor approximation, we also need a bandlimited envelope function with (at least) $N-1$ continuous derivatives and a  $N^{\text{th}}$ derivative of bounded variation. To this end we here use the function
\begin{equation}
e(t) = \text{sinc} ^{N+1} \left( \frac{t}{(N+1)\delta} \right) \quad \text{with} \quad \delta \geq 1,
\label{eq:Envelope}
\end{equation}
where $\text{sinc}(x)=\sin(\pi x)/(\pi x)$. Through $N$ successive convolutions of the Fourier transform of the sinc function (which is equal to 1 for $|\omega| < \pi$ and zero otherwise) with itself, it is easy to show that this function satisfies our requirement plus it has a spectrum that is supported in $[-\pi/\delta,\pi/\delta]\subseteq[-\pi,\pi]$. The parameter $\delta$ (dilation factor) is used to control the flatness of $e(t)$ around $t=0$ and hence the accuracy $\epsilon_2>0$ with which $f_{\text{BL}}(t) = p_N(t)e(t)$ approximates $p_N(t)$ in the interval $(-a,a)$. Obviously, a high accuracy (a small $\epsilon_2) $ requires $a \ll (N+1)\delta$, so that $|e(t)-1| \ll 1$ for $|t|<a$. At the same time, we require the suboscillation (here the constant behaviour) to last at least one period of the minimum frequency contained in $f_\text{BP}(t)$, $\omega_\text{min} = \Omega - \pi/\delta$. The two conditions are combined in
\begin{equation}
\pi \left( \Omega - \pi/\delta\right)^{-1} \leq a \ll (N+1)\delta.
\label{eq:Conditions}
\end{equation}
Figure \ref{fig:figure_1} shows a specific example of such a suboscillatory function for the parameters $a=1$, $N=19$, $\Omega = 2\pi$ and $\delta=4$. Part (a) of this figure shows the real and imaginary parts of $f_\text{BP}(t)$, verifying that the function approximates the constant function $s(t)=1 + i0$ in the interval $(-1,1)$. To quantify the accuracy of the approximation, the deviation from the desired profile is also shown in logarithmic scale and it is seen that, under the specific selection of parameters, $|f_\text{BP}(t) - s(t)|<10^{-2}$ for $t \in (-1,1)$. Notice also how the deviation increases rapidly as soon as $t$ departs from the interval $(-1,1)$. The profile of the minimum frequency in the spectrum of $f_\text{BP}(t)$, namely $\omega_\text{min} = 7\pi/4$, is also plotted in the same figure to give a sense of the extent of the suboscillatory behaviour. The interval $(-1,1)$ in which $f_\text{BP}(t)$ approximates the constant function $s(t)=1$ is equal to $1.75$ periods of this minimum frequency.

An extended view of the function is shown in part (b), where the large difference in scale (24 orders of magnitude) is noted between the values of the function in the region of ``normal" oscillations and in the suboscillatory interval $(-1,1)$, an inevitable feature of both superoscillatory and suboscillatory functions \cite{Ferreira2006FasterThanNyquist}. Notice that the ``normal" oscillations of $f_{\text{BP}}(t)$ are due to the modulation factor $e^{i 2 \pi t}$ but are too dense to be discerned in the extended interval of part (b). 

Finally, part (c) depicts the spectrum of $f_\text{BP}(t)$ which is strictly zero outside $[7\pi/4,9\pi/4]=[1.75\pi,2.25\pi]$. The spectrum has discontinuities which are depicted as vertical segments. The discontinuities are expected since $f_\text{BP}(t)$ results from the product of a $19^{\text{th}}$ order polynomial with the envelope $\text{sinc} ^{20} \left( t / 80  \right)$ whose Fourier transform has $18$ continuous derivatives and a discontinuous (but bounded) $19^{\text{th}}$ derivative. By selecting  envelope functions $e(t)$ with smoother spectra, say with $N+m$ continuous derivatives  $(m \geq 0)$ (which of course implies that $e(t)$ decays at least as $O(|t|^{-(N+m+2)})$ as $|t| \to \infty$), we obtain suboscillatory functions with correspondingly smoother spectra of $m$ continuous derivatives.

\section{Conclusion}
We have shown through simple arguments that it is possible to obtain bandpass functions that approximate a given analytic function with arbitrarily small error over an arbitrarily long, yet finite, interval. If the function being approximated oscillates in the interval with local frequencies that are smaller that the minimum frequency component of the bandpass function, the latter is called suboscillatory. We illustrated the method with the most extreme case of suboscillations which is that of a function that stays (almost) constant over an arbitrarily long interval without containing a dc component $(\omega=0)$ in its spectrum.

The present method was based on the existence of bandlimited functions that approximate a given polynomial with arbitrary accuracy over an interval, which we reported in \cite{Chremmos_SOPolynomial} with emphasis on the design of superoscillations. The polynomial can be generalized to an arbitrary analytic function, as done in this work for bandpass functions. Hence, both bandlimited and bandpass functions can be made to approximate arbitrary analytic functions over an interval. In fact, the method that we described here allows us to reach a more general statement:\\

\textit{Given an analytic function $s(t)$ of the real variable $t$, an interval $(a,b)$, $\epsilon>0$, and a frequency band $[\omega_1,  \omega_2]$ (that may or may not contain zero), there exists a (at least one) function $f(t)$ whose spectrum is supported in $[\omega_1, \omega_2]$ and which satisfies $|f(t) - s(t)| < \epsilon$ for $t \in (a,b)$.}\\

Such a function $f(t)$ can be constructed with the present method, where the frequency band $[\Omega - \pi$, $\Omega + \pi]$ is identified as $[\omega_1$, $\omega_2]$. Now, if the band of frequencies contains zero (namely if $\omega_1 \omega_2 < 0$) and $s(t)$ oscillates in $(a,b)$ with local frequencies $|\omega_{l}| > \text{max}\{|\omega_1|, \omega_2\}$, the function $f(t)$ is called superoscillatory, while if the band of frequencies does not contain zero (namely if $\omega_1 \omega_2 > 0$) and $s(t)$ oscillates in $(a,b)$ with local frequencies $|\omega_{l}| < \text{min}\{|\omega_1|, |\omega_2|\}$, the function $f(t)$ is called suboscillatory. Such a unified view of superoscillations and suboscillations is reported here for the first time to our knowledge.

\bibliography{mypapers}
\bibliographystyle{ieeetr}

\end{document}